\documentstyle[prl,aps,twocolumn,epsfig,amssymb]{revtex}
\begin{document}
\draft

\title{Understanding depletion forces beyond entropy}

\author{C. Bechinger, D. Rudhardt, and P. Leiderer}
\address{Fakult\"at f\"ur Physik, Universit\"at Konstanz, D-78457 Konstanz, Germany}
\author{R. Roth and S. Dietrich}
\address{Fachbereich Physik, Bergische Universit\"at Wuppertal, D-42097
  Wuppertal, Germany}

\maketitle
\begin{abstract}
The effective interaction energy of a colloidal sphere in a suspension containing
small amounts of non-ionic polymers and a flat glass surface has been measured and
calculated using total internal reflection microscopy (TIRM) and a novel approach
within density functional theory (DFT), respectively. Quantitative agreement
between experiment and theory demonstrates that the resulting repulsive part of the
depletion forces cannot be interpreted entirely in terms of entropic arguments but 
that particularly at small distances ($\lesssim$ 100 nm) attractive dispersion 
forces have to be taken into account.

\end{abstract}


\narrowtext
The stability of systems like polydisperse mixtures of colloids, natural 
rubbers, micelles or polymer coils is known to be strongly influenced by
depletion forces. In addition, such forces may also play an important role
in biological systems, e.g., in promoting the aggregation of red blood cells
\cite{Mayer45,Janzen89}. The understanding of these forces being responsible for
spontaneous structure formation, phase separation or flocculation in such systems
is highly demanding both from the experimental and the theoretical point of view. 
A first approximate explanation of depletion forces was given by Asakura and Oosawa
(AO), who recognized that the presence of small hard spheres (referred to as 
macromolecules in the following) can mediate effective forces between two larger 
objects if their distance is sufficiently small \cite{Asakura54}. This can be
easily understood by considering, e.g., a hard sphere of radius $R$ suspended in a 
hard-sphere fluid of macromolecules with radius $r$ and bulk number density 
$\rho_b$ in front of a wall (or another big sphere) at distance $z$, measured
between the surfaces of the wall and the spheres. For $z < 2 r$, the macromolecules
are expelled from a region of excluded volume overlap, i.e., their density is
depleted at that hemisphere of the large particle facing the wall compared to the 
opposite 
side which faces the bulk liquid. Such density gradients give rise to an effective 
osmotic pressure, causing the big sphere to be pushed towards the wall thereby
allowing the entropy of the macromolecules to increase. Accordingly, these 
depletion forces result from an asymmetric density distribution of the
macromolecules around the big sphere. Quantitative calculations of depletion forces
require the knowledge of the density distribution $\rho({\bf r})$ of the 
macromolecules. Within the crudest approximation $\rho({\bf r})$ is considered to
be constant (corresponding to the above mentioned AO-approximation), so that
depletion forces can be easily calculated in terms of excluded volume arguments and
are predicted to be entirely attractive for $z < 2 r$. For sufficiently low 
macromolecule densities this approximation is in agreement with experimental
results \cite{Ohshima97,Rudhardt98}. If, however, at high concentrations, the 
structural correlation effects in the macromolecular liquid become important,
$\rho({\bf r})$ displays an oscillatory behavior at small values of $z$ generating 
repulsive parts of the depletion interactions as observed recently \cite{Crocker99}.
By means of virial coefficient expansions within the Derjaguin approximation
\cite{Mao95} and by other thechniques in \cite{Goetzelmann98} it has been found
that the repulsive interaction strength strongly increases for large values of 
$R/r$.

So far the interpretation of the corresponding experiments has been based on
assuming hard sphere macromolecule and hard wall interactions which is, however, 
not always justified under experimental conditions. This is most evident in the
case of highly charged macromolecules where even at volume fractions 
$\rho_b 4 \pi r^3/3 \lesssim 0.02$ electrostatic interactions lead to long-range 
repulsive and short-range attractive parts in the depletion potential between a 
sphere and the wall \cite{Sober95,Sharma96}. But even for systems of uncharged
macromolecules or in the high salt limit, the model of hard spheres {\em and} a
hard wall is not applicable within the range of attractive dispersion forces.

Here we present measurements of the depletion potential of a single colloidal
polystyrene (PS) sphere in front of a flat glass surface in the presence of
non-ionic macromolecules for various values of the macromolecule density $\rho_b$. 
In addition we have calculated the depletion potential within DFT for a hard-sphere
system in which the dispersion and electrostatic interactions between the 
macromolecules and the wall have been taken into account. We find clear evidence 
for the occurrence of attractive and repulsive interaction energies both in our 
experiment and in our calculations. The latter demonstrate that  dispersion forces
considerably modify the depletion potential below 100 nm ($\simeq r$). This
substantiates the hypothesis \cite{Dinsmore99} that in general for depletion forces
both entropic and energetic contributions are important.

We have used TIRM \cite{TIRM} to measure the potential energy of a 
colloidal sphere close to a glass surface. In this technique, a laser beam is 
reflected from a solid-fluid interface slightly above the critical angle of total
reflection thus generating an evanescent wave in the fluid. A colloidal sphere 
located sufficiently close to the interface scatters the evanescent wave with an
intensity which decays exponentially with increasing particle-interface distance. 
Thus the scattered intensity, which fluctuates owing to Brownian motion, determines
sensitively and instantaneously the separation distance $z$. In order to obtain the
spatial dependence of the potential energy of the particle one has to measure the
separation distances sampled by the colloidal sphere for a statistically long
period of time. From this, the probability of finding the particle at any
separation distance can be calculated which is related to the potential energy via
the Boltzmann distribution. If the penetration depth of the evanescent wave is
chosen to be of the order of several hundred nanometers, this technique allows one
to determine precisely effective particle-wall interaction potentials close to an
interface.

Monodisperse PS particles of radius $R$=5 $\mu$m were stabilized with sulfate
surface groups which dissociate in contact with water and thus cause the particles 
to be negatively charged. Owing to the large weight of the PS spheres, it was 
necessary to increase their buoyancy to facilitate large enough thermally driven 
vertical excursions of the particles during the measurements. Therefore the 
particles were suspended in a mixture of deuterated water and normal water which 
caused a reduction of the weight by a factor of 8.5. As macromolecules we chose 
poly(ethylene oxide) (PEO) which had been successfully employed in other studies of
depletion effects before \cite{Ohshima97,Rudhardt98,Rudhardt99}. PEO is a non-ionic
water-soluble polymer which forms polymer coils below a critical density $\rho_c$.
For the molecular weight which was used in our experiments ($M_W=10^6$) the 
radius of gyration $r_G$ has been determined by means of static light scattering to
be $r_G$=67.7 nm \cite{Devanand91}. The maximum polymer number density used in 
our experiments was 23.4 $\mu$m$^{-3}$ which is smaller than $\rho_c$ by more than
a factor of 30. Accordingly in the present context the polymer coils can be
considered as spherical macromolecules to be modeled in our calculations as
hard spheres ($\rho_b \ll \rho_c$) with an effective radius $r$=107 nm 
\cite{radius}.

The sample cell was composed of two silica flats separated by a 1 mm thick O-ring
and was connected to a closed circuit system containing an ion exchange resin and a
conductivity probe. This setup allowed us to control precisely the ionic strength
of the suspension. The evanescent field was generated by a HeNe-laser beam 
($\lambda$ = 633 nm, P = 5 mW) which was coupled to the bottom plate of the cell
under total internal reflection conditions by means of a glass prism. The 
penetration depth was chosen to be 400 nm. Highly diluted suspensions with PS 
number densities below 0.5 mm$^{-3}$ have been used to ensure that only a single PS
particle was in the field of view and contributed to the scattering signal. The
scattered intensity of the PS sphere was collected by a microscope objective and
focused onto a photomultiplier tube. In order to obtain sufficient data and to 
minimize statistical errors the scattered intensity of a particle was sampled over 
at least 1000 s at a sampling rate of 50 Hz.

In the absence of depletion forces, i.e., without adding macromolecules, the
potential of a negatively charged PS particle above a glass plate is composed of
two parts. Towards smaller distances the potential increases exponentially owing to
the electrostatic interaction between the sphere and the fused silica plate, which 
is also negatively charged when in contact with water. For larger distances the 
potential is completely dominated by gravity. It has been shown theoretically and 
experimentally that under these conditions the external potential is given by 
\cite{TIRM}
\begin{equation}\label{potential}
\beta \Phi_{ext}^{PS}(z) = B \exp(-\kappa z) + \beta~ G z
\end{equation}
where $B$ depends on the sphere and the glass plate surface potentials, $\kappa$ is
the inverse Debye screening length related to the ionic strength of the suspension,
and $\beta=(k_B T)^{-1}$.  $G = (\bar \rho_{PS}-\bar \rho_W) V g$ is the weight of
the particle of volume $V$ suspended in water, $\bar \rho_{PS}$ and $\bar \rho_W$ 
are the mass densities of PS and water, respectively, and $g$ is the acceleration 
of gravity. The values for $V$ and $\bar \rho_{PS}$ were provided by the 
manufacturer.
 
\begin{figure}
\centering\epsfig{file=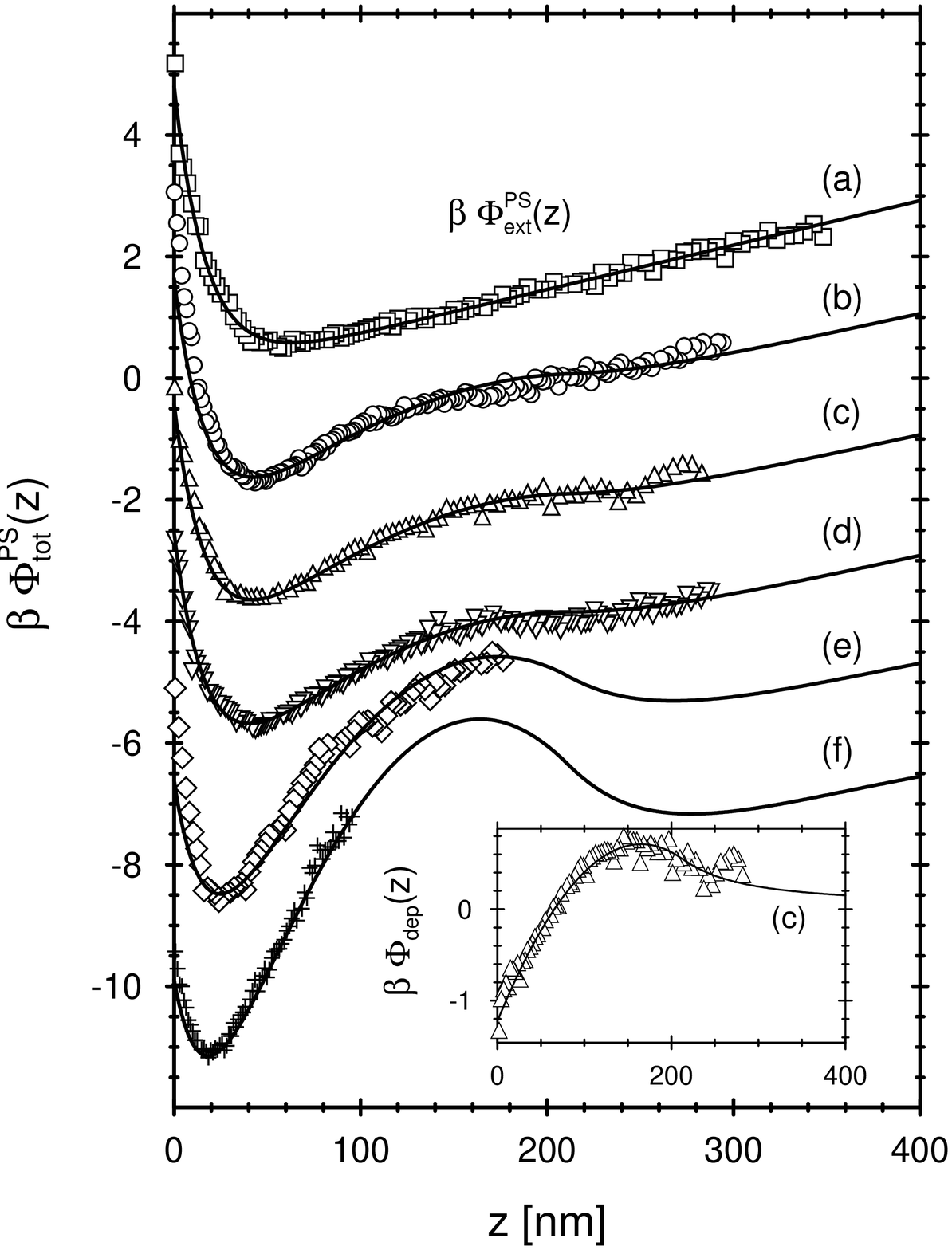,bbllx=0,bblly=60,bburx=540,bbury=770,width=13.5cm}
\vspace{0.5cm}
\caption{\label{fig:exp} Comparison of the effective potential 
$\Phi_{tot}^{PS}(z)=\Phi_{ext}^{PS}(z)+\Phi_{dep}(z)$ acting on the big PS particle 
between theory (solid lines) and experiment (symbols) for various number densities
of the macromolecules: (a) $\rho_b^{exp}$ = 0, (b) 4.1 $\mu$m$^{-3}$ , (c) 6.3
$\mu$m$^{-3}$, (d) 8.7 $\mu$m$^{-3}$, (e) 16.4 $\mu$m$^{-3}$, and (f) 23.4
$\mu$m$^{-3}$. For reasons of clarity the curves have been separated from each 
other vertically by 4 units.}
\end{figure}

This potential is shown in Fig.~\ref{fig:exp}(a) as measured ($\square$). The PS 
sphere probes a range of about 300 nm from the surface during our measuring time.
Due to the large weight of the PS sphere with $R$ = 5 $\mu$m, we are very sensitive
to changes in the sphere-wall potential below $z$ = 150 nm; for smaller spheres this
potential is dominated by the electrostatic repulsion \cite{Rudhardt98}. The solid 
line in Fig.~\ref{fig:exp}(a) is the fit curve according to Eq.~(\ref{potential}) 
with the two fit parameters $B = 4.81$ and $\kappa^{-1}$ = 17 nm (for $T$=298 K), 
the latter being in agreement with the electrical conductivity measurement of the
colloidal circuit. This demonstrates that in Eq.~(\ref{potential}) attractive
dispersion forces are negligible.

However, for the macromolecules dispersion and electrostatic interactions between 
them and the glass plate have to be taken into account. The latter originate from 
the fact that the charged glass plate induces a dipole moment in the neutral
macromolecules which interacts with the electric field resulting in a repulsion
due to the much smaller static index of refraction $\varepsilon_1$ of PEO than that
of water ($\varepsilon=81$). This leads to
\begin{equation}\label{external}
\Phi_{ext}^{PEO}(z) = C \exp(-2 \kappa (z+r))) - A \left(\frac{r}{z+r}\right)^3
\end{equation}
with the Hamaker constant $A = 5\times 10^{-20}$ J \cite{Hamaker} and
\begin{equation}
C = 12 \pi \left(\kappa r \cosh(\kappa r) - \sinh(\kappa r)\right)
\frac{\varepsilon~ \varepsilon_0 (\varepsilon - \varepsilon_1)} 
{\varepsilon_1+2 \varepsilon}\frac{E_0^2}{\kappa^3}
\end{equation}
with $\varepsilon_1\approx 5$ \cite{epsilon}, the surface electric potential of the 
glass plate $E_0\approx 50$ mV \cite{E0}, and $\varepsilon_0$ the permittivity
of vacuum. Estimates indicate that for the densities $\rho_b$ considered here
the attractive dispersion forces among the macromolecules and between them and the
PS spheres are not important. The agreement, which we find between our experimental
data and our calculations without considering the latter forces, supports this
approximation a posteriori.

When PEO is added to the system the presence of depletion forces generates modified
effective potentials acting on the PS sphere. Figure~\ref{fig:exp} shows the 
measured potentials (symbols) $\Phi_{tot}^{PS}(z)=\Phi_{ext}^{PS}(z)+\Phi_{dep}(z)$ 
when different amounts of PEO are added to the suspension. Assuming a 
homogeneous distribution of PEO over the whole circuit, from the experimentally 
determined weights, the PEO number densities are obtained as (b) $\rho_b^{exp}$ = 
4.1 $\mu$m$^{-3}$, (c) 6.3 $\mu$m$^{-3}$, (d) 8.7 $\mu$m$^{-3}$, (e) 16.4 
$\mu$m$^{-3}$, and (f) 23.4 $\mu$m$^{-3}$, respectively. With increasing PEO 
concentration the potential deepens and the position of the minimum considerably 
shifts to smaller values of $z$ in agreement with previous measurements 
\cite{Rudhardt98}. Additionally, however, we find a bump in the potential around z 
= 150 nm which increases with increasing PEO densities. In order to highlight this 
feature we have subtracted the electrostatic and gravitational contribution 
(Eq.~(\ref{potential})) from  Fig.~\ref{fig:exp}(c) and present this difference in
the inset of Fig.~\ref{fig:exp} as measured ($\triangle$) and as calculated (solid
line). Within $\Phi_{dep}$ the bump turns into a pronounced maximum with a height
of approximately 0.75 $k_B T$ which gives rise to a repulsive depletion force of
the PS sphere upon approaching the wall before it finally becomes attractive for
$z$ smaller than 150 nm. 

\begin{figure}
\centering\epsfig{file=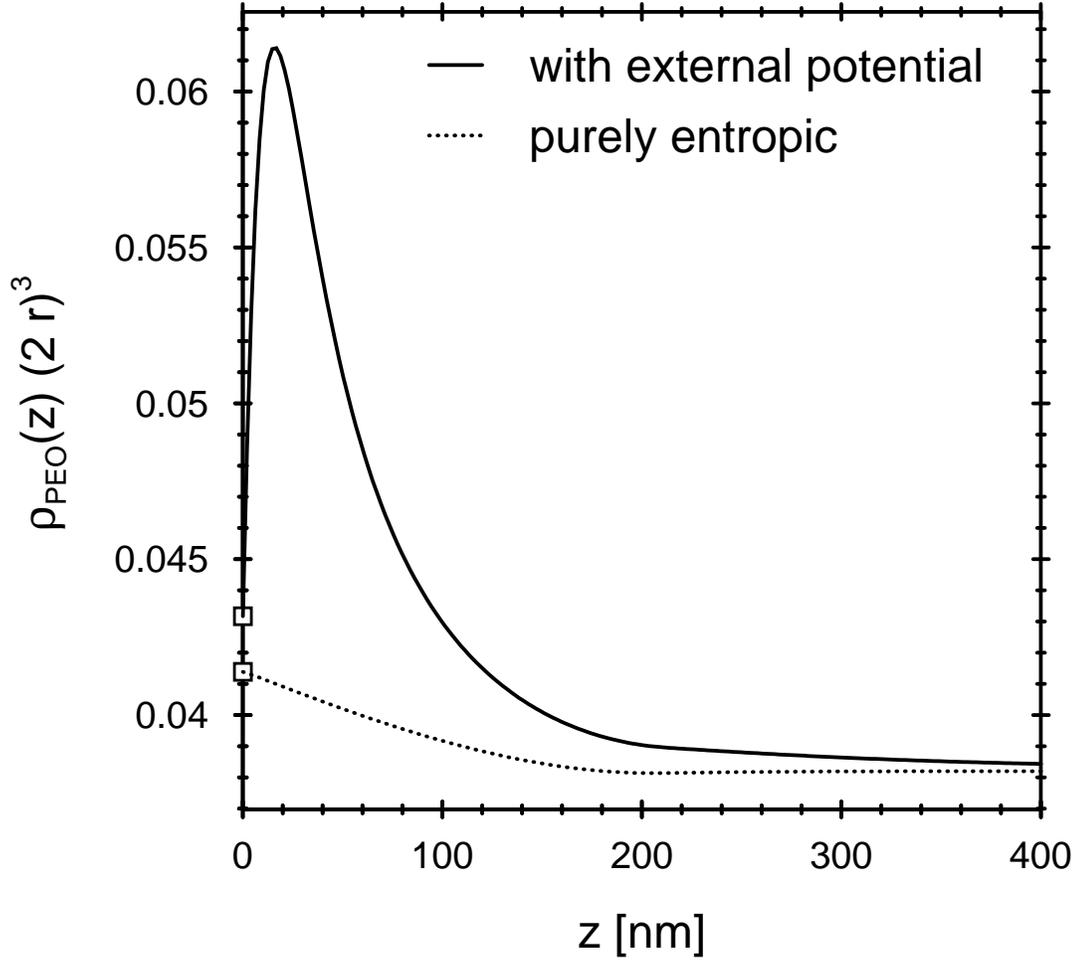,bbllx=0,bblly=60,bburx=540,bbury=770,
width=13.5cm}
\vspace{0.5cm}
\caption{\label{fig:prof} The number density profile $\rho_{PEO}(z)$ of the 
macromolecules in the absence of a PS sphere for the wall potential
$\Phi_{ext}^{PEO}(z)$ given by Eq.~\ref{external} (solid line) and for a hard wall
(dotted line); the theoretical value $\rho_b^{th}$ = 4.1 $\mu$m$^{-3}$ corresponds 
to Fig.~\protect\ref{fig:exp}(c) and $\rho_{PEO}(z=0) (2 r)^3$ = 0.043.}
\end{figure}

Our theoretical calculations for $\Phi_{tot}^{PS}(z)$ are based on a novel and 
versatile DFT approach \cite{Goetz99}. It has been shown that in the limit of 
present interest, i.e., when the density of PS goes to zero, the depletion 
potential $\Phi_{dep}(z)$ can be determined in two steps {\em solely} from the 
density profile of the macromolecules $\rho_{PEO}(z)$ {\em without} PS sphere and
from the geometrical shape of the PS particle, i.e., its radius $R$ \cite{Goetz99}. 
First, the density profiles of the polymers $\rho_{PEO}(z)$ subjected to the
external potential given by Eq.~(\ref{external}) have been obtained by freely
minimizing the Rosenfeld functional \cite{Rosenfeld89}. With all parameters in
the external potential and the effective hard sphere radius $r$ of the
macromolecules fixed, the only adjustable parameter in the calculation is the PEO
bulk density. In Fig.~\ref{fig:prof} we show a PEO density profile for $\rho_b$=4.1
$\mu$m$^{-3}$ (solid line). The density profile reflects the short ranged 
electrostatic repulsion at very small distances and the dispersion attraction 
leading to a pronounced maximum more than $50\%$ above the bulk value. The
importance of $\Phi_{ext}^{PEO}$ for bringing about this structure is underscored 
by the dotted line in Fig.~\ref{fig:prof} which corresponds to 
$\Phi_{ext}^{PEO}\equiv 0$. Second, from such PEO density profiles we have 
calculated the effective potentials shown as solid lines in 
Fig.~\ref{fig:exp}(b)-(f). As mentioned above, the only adjustable parameter in the
calculations is $\rho_b$ which has been varied until best agreement with the
experimental data has been obtained leading to the theoretical values (b) 
$\rho_b^{th}$ =3.9 $\mu$m$^{-3}$, (c) 4.1 $\mu$m$^{-3}$, (d) 4.5 $\mu$m$^{-3}$,
(e) 11.3 $\mu$m$^{-3}$, and (f) 15.5 $\mu$m$^{-3}$, respectively. We find very
good agreement with the experimental potential data and in particular the emerging
barrier for increasing $\rho_b$ is reproduced. The values of $\rho_b^{th}$ agree
within a factor of about $1.5$ with $\rho_b^{exp}$. A possible reason for these
differences could be due to  lateral density gradients within the experimental 
circuit not captured by the aforementioned global determination of $\rho_b^{exp}$.

\begin{figure}
\centering\epsfig{file=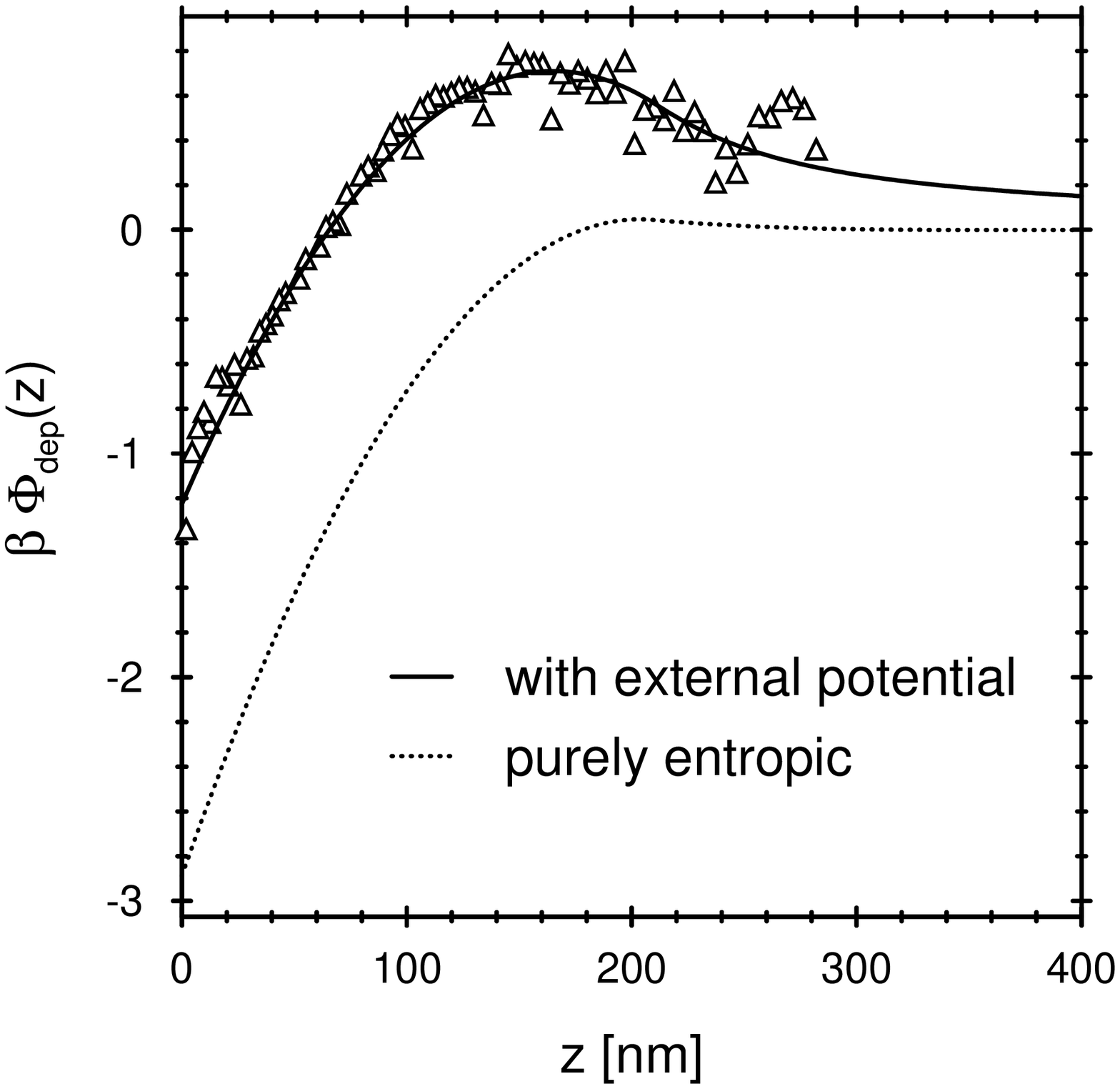,bbllx=0,bblly=60,bburx=540,bbury=770,
width=13.5cm}
\vspace{0.5cm}
\caption{\label{fig:pot} The depletion potentials $\Phi_{dep}(z)$ between the wall 
and a big PS sphere calculated from the corresponding PEO density profiles
$\rho_{PEO}(z)$ shown in Fig.\protect\ref{fig:prof}. $\triangle$ denote the 
corresponding experimental data.}

\end{figure}

We emphasize that the potential barriers in Fig.~\ref{fig:exp} cannot be explained
in terms of a hard-sphere and hard-wall model, i.e., by purely entropic arguments.
This conclusion is demonstrated in Fig.~\ref{fig:pot} by the discrepancy between
the actual depletion potential and the one close to a hard wall.

Another important conclusion which can be drawn from our calculations is that,
although the dispersion attraction seems to be mainly responsible for the potential
barrier, the soft electrostatic repulsion plays an important role as well. By
changing the salt concentration and hence the inverse screening length, one can 
experimentally tune the range of the electrostatic repulsion and thus modify the 
potential barrier. This explains why in an earlier experiment with a lower salt
concentration no repulsive depletion forces were found \cite{Rudhardt98}.

So far, we have demonstrated that DFT allows one to calculate accurately the
depletion force on a suspended sphere from the density profile of the dissolved
macromolecules in the {\em absence} of the big sphere. For small macromolecule
densities this relationship can be inverted: measuring the depletion potential of a
large sphere may be used to determine the {\em undisturbed} macromolecule density
profile. This is a remarkable result, because the big sphere considerably modifies
the density distribution of macromolecules. Thus depletion force measurements can be
used as a tool for determining undisturbed solvent properties.

In summary, we have studied the depletion potential of a single polystyrene
sphere in a fluid of non-ionic macromolecules in front of a wall. The potential 
exhibits both attractive as well as repulsive parts which are not of purely
entropic origin but are largely dominated by the dispersion forces between the
uncharged macromolecules and the wall. These results are in good quantitative
agreement with density functional theory calculations carried out specificly for 
this system. 

C.B., D.R., and P.L acknowledge financial support from the Deutsche
Forschungsgemeinschaft through SFB 513.

\onecolumn

\begin{references}
\bibitem{Mayer45} K. Mayer, E. Hahnel, and R.R. Feiner, Proc. Roy. Exp. Biol. Med.
{\bf 58}, 36 (1945).
\bibitem{Janzen89} J. Janzen and D.E. Brooks, Clinical Hemorheology {\bf 9}, 695 
(1989).
\bibitem{Asakura54} S. Asakura and F. Oosawa, J. Chem. Phys. {\bf 22}, 1255 (1954).
\bibitem{Ohshima97} Y.N. Ohshima, et al., Phys. Rev. Lett. {\bf 78}, 3963 (1997).
\bibitem{Rudhardt98} D. Rudhardt, C. Bechinger, and P. Leiderer, Phys. Rev. Lett.
{\bf 81}, 1330 (1998).
\bibitem{Crocker99} J.C. Crocker, J.A. Matteo, A.D. Dinsmore, and A.G. Yodh,
Phys. Rev. Lett. {\bf 82}, 1999.
\bibitem{Mao95} Y. Mao, M.E. Cates, and H.N.W. Lekkerkerker, Physica A
{\bf 222}, 10 (1995).
\bibitem{Goetzelmann98} B. G\"otzelmann, R. Evans, and S. Dietrich, Phys. Rev.
E {\bf 57}, 6785 (1998).
\bibitem{Sober95} D.L. Sober and J.Y. Walz, Langmuir {\bf 11}, 2352 (1995).
\bibitem{Sharma96} A. Sharma and J.Y. Walz, J. Chem. Soc., Faraday Trans. {\bf 92},
4997 (1996).
\bibitem{Dinsmore99} A.D. Dinsmore and A.G. Yodh, Langmuir {\bf 15}, 314 (1999).
\bibitem{TIRM} D.C. Prieve, F. Luo, and F. Lanni, Faraday Discuss. Chem. Soc. 
{\bf 83}, 297 (1987); D. Haughey and J.C. Earnshaw, Colloids Surf. A {\bf 136}, 217 
(1998).
\bibitem{Rudhardt99} D. Rudhardt, C. Bechinger, and P. Leiderer, Prog. Colloid 
Polym. Sci. {\bf 112}, 163 (1999).
\bibitem{Devanand91} K. Devanand and J.C. Selser, Macromolecules {\bf 24}, 5943
(1991).
\bibitem{radius} If due to thermal averaging  PEO is considered as a homogeneous 
hard sphere with radius $r$, this value is obtained from the radius of gyration by 
$r=\sqrt{5/2}~ r_G$ which yields $r$ = 107 nm for $r_G$ = 67.7 nm (see D.F. Evans 
and H. Wennerstr{\"o}m, {\em The Colloidal Domain} (VCH, New York, 1994), p. 292).
\bibitem{Hamaker} J. Israelachvili, {\em Intermolecular \& Surface Forces} 
(Academic, London, 1992).
\bibitem{epsilon} No accurate values for $\varepsilon_1$ were found in the 
literature. Therefore we varied $\varepsilon_1$ between $3$ and $10$ which did not 
result in any visibly changes of either the density profiles or the depletion
potentials.
\bibitem{E0} S. Tanimoto, H. Matsuoka, H. Yamauchi, and H. Yamaoka, Colloid 
Polym. Sci. {\bf 277}, 130 (1999).
\bibitem{Goetz99} B. G\"otzelmann, R. Roth, S. Dietrich, M. Dijkstra, and R. Evans,
Europhys. Lett. {\bf 47}, 398 (1999).
\bibitem{Rosenfeld89} Y. Rosenfeld, Phys. Rev. Lett. {\bf 69}, 980 (1989).
\end{references}
\end{document}